\documentclass[epjST]{svjour}
\usepackage{graphics,amssymb,amsmath,bm}

\newcommand{\rf}[1]{(\ref{#1})}
\begin{document}
\title{Periodic solutions to a mean-field model for electrocortical activity}
\author{Lennaert van Veen\thanks{\email{Lennaert.vanVeen@uoit.ca}}\fnmsep\inst{1} \and
  Kevin R. Green\inst{1,2}}
\institute{University of Ontario Institute of Technology, Oshawa, Ontario, Canada \and 
  INRIA-Nancy Grand Est, team NEUROSYS, Villers-l\`es-Nancy, France}
\abstract{
We consider a continuum model of electrical signals in the human cortex, which takes the form of a 
system of semilinear, hyperbolic partial differential equations for the inhibitory and excitatory membrane
potentials and the synaptic inputs. The coupling of these components is represented by 
sigmoidal and quadratic nonlinearities. We consider these equations on a square domain with
periodic boundary conditions, in the vicinity of the primary transition from a stable equilibrium
to time-periodic motion through an equivariant Hopf bifurcation. We compute part of a family of 
standing wave solutions, emanating from this point. 
} 
\maketitle
\section{Introduction}
\label{intro}
A salient, if ill-understood, feature of the human cortex is that it produces electrical fields
that can be coherent on a length scale of millimetres to centimetres and a time scale of milliseconds
to seconds. The biophysical origin of such fields can be explained accurately in terms of the physiology
of the neurons, densely packed in the cortical layers \cite{Nunez2006}. When hundreds of thousands of neurons,
spread out over a few square centimetres of cortical tissue, fire synchronously, they can produce an
electrical field strong enough to measure on top of the scalp. This measurement is called the 
electroencephalograph (EEG), and has been used over the past century to uncover coherence in electrocortical
activity. It is non-invasive and cheap compared to more recently introduced imaging techniques, such as functional Magnetic
Resonance Imaging (fMRI). Although it does not have great spatial resolution, due to the spreading of signals
in the scalp, its temporal resolution is high, and
it has been used in countless experiments probing the
processing of sensory stimuli, the generation of pathological states such as
seizures and aspects of cognition.

From this body of experimental work, we have learned a lot about the phenomenology of
electrical activity in the cortex. For instance, it is known that fluctuations at different
frequencies play different roles in the processing of stimuli. Activity in the alpha band (8--13Hz)
appears to indicate ``readiness'', a state the cortex can easily move to and from during processing
\cite{Liley2009}, while gamma-band oscillations (30-100Hz) are thought to facilitate the integration of 
networks of neurons into coherent activity (e.g. \cite{rolls2012} and references therein).
Experimental work has also enabled the use of the EEG for diagnostic purposes,
for instance when deciding on treatment for epilepsy or to assess the level of consciousness of
patients under anaesthesia.

In contrast, questions pertaining to the organisation and dynamics of coherent electrical activity,
as well as to its role in high-level brain function, are much harder to answer. Fundamental 
questions, such as whether alpha-band activity is generated by the cortex itself, or is a result
of interaction with deeper brain structures like the thalamus, are, as yet, unanswered. To address such
questions, phenomenology
alone does not suffice; we need to integrate observations with predictive models.

The two most common approaches to modelling electrocortical activity are based on the simultaneous 
simulation of many individual, coupled neurons and on coarse-grained simulations of
ensembles of neurons. In the former approach, a large, densely connected network of elementary
neuronal units is time-stepped, and the macroscopic electrical potential can be reconstructed
by summing over the individual outputs. Using state-of-the-art computing facilities, as many
as one million units, tied together by half a billion connections, have been simulated
and tuned to mimic output  measured {\em in vitro} in small mammals \cite{izhi2008}.
In the latter approach,  mean properties of ensembles of
neurons are modeled directly. The dependent variables can then be thought of as locally 
averaged over an area of about one square millimetre. This is often labeled the {\em mean field}
approach. Under simplifying assumptions on the stochastic nature of individual neuron dynamics
and on the density of their connections, mean field models can be derived from
``many neuron models'' as a kind of thermodynamic limit \cite{deco2008}.

Mean-field models, in turn, come in different flavours. One important distinction is that between
convolution-based models and conductance-based models \cite{pinotsis2013}. In the former, the
time derivatives of the mean fields are given directly by convolutions of these fields with kernels
that specify the connectivity. A classical example is the Wilson-Cowan model \cite{wilson1973}, in which
the fields represent the fractions of excitatory and inhibitory neurons that are firing.
In the latter, the fields typically represent the mean excitatory and inhibitory membrane potential
and the conductances associated with the synaptic coupling between them.
Within the class of conductance-based models, there are several
different types of models, distinguished by the way the input into a local cluster of neurons
is computed from the global fields. This relation is most naturally expressed in terms of an integral over inputs,
originating anywhere on the cortical surface, with a kernel that specifies the connection strength
and delay time as a function of the distance to the source. This formulation results in
a system of nonlinear integro-differential equations with distributed delays. Both the
theoretical and the numerical treatment of such models is very challenging. Although much
progress has been made in the analysis of simplified models, for instance with only a single
neuronal population or only a single spatial dimension (see, e.g. \cite{meijer2014,bressloff2012}),
a direct analysis of full-fledged models of this kind is not yet feasible. 

A great simplification is achieved by approximating the delay integro-differential equations
by damped wave equations, a step explained in detail by Liley {\sl et al.} \cite{liley2002}. The result
is a system of nonlinear Partial Differential Equations (PDEs), which can be investigated in terms
of the theory of continuous dynamical systems, not dissimilar to, for instance, the Navier-Stokes equations
for fluid motion. This analysis is still quite challenging for a number of reasons, including
\begin{itemize}
\item the strong nonlinearities, in particular the sigmoidal functions that relate the change in conductance
to pre-synaptic inputs;
\item the large number of parameters, upon which the dynamics can depend very sensitively;
\item the difficulty in determining physiologically acceptable boundary conditions and
\item the presence of continuous and discrete symmetries, depending on  the choice of boundary conditions.
\end{itemize}
For this reason, a lot of work on PDE-based mean field models focused on parameter regimes where
a stable equilibrium state exists (e.g. \cite{bojak2005,pinotsis2013}). In this work, the
response of the mean-field models to noisy external input was investigated, and it was found that the models
can reproduce realistic temporal spectra for the membrane potentials, which can be compared
to EEG spectra.

In the current work, we attempt to study the dynamical repertoire of a PDE-based mean field model
beyond the primary instability of the equilibrium state. We
focus on Liley's model, which describes both inhibitory and
excitatory populations and their short range and long range connections \cite{liley2002}. In total, there are fourteen fields 
that satisfy either local equations or hyperbolic, semilinear PDEs. In earlier work, we reported on the software package
MFM for the simulation of this model \cite{green2013}. Here, we present the computation of a spatially
inhomogeneous, time-periodic solution, which takes the form of a standing wave, or a {\em standing square} in the
language of equivariant bifurcation theory.
As far as we are aware, this is the first such computation in a PDE-based mean field model with both excitatory
and inhibitory populations and two spatial dimensions. We hypothesize that, when continued in an external
input parameter, this periodic orbit will exhibit the formation of ``hot spots'' of gamma-band activity seen in
the highly nonlinear regime of this model \cite{bojak2007}. 

\section{Brief description of the model}
\label{sec:model}
The dependent variables of Liley's model are the mean inhibitory and excitatory membrane potential,
$h_i$ and $h_e$, the four mean synaptic inputs, originating from either population 
and connecting to either, $I_{ee}$, $I_{ei}$, $I_{ie}$ and $I_{ii}$, and the excitatory
axonal activity in long-range fibres, connecting to either population, $\phi_{ee}$ and $\phi_{ei}$.
In this context, {\em long-range} means over one millimetre in length. Such connections originating from inhibitory neurons 
are excluded, based on {\em in vitro} measurements indicating that they are extremely rare (e.g. \cite{stepanyants2009}).

The model equations are
\begin{subequations}
\begin{equation} \label{eq:Liley1}
\tau_k \frac{\partial h_k(\vec{x},t)}{\partial t} = h^r_k - h_k(\vec{x},t) 
  + \frac{h^{eq}_{ek}-h_k(\vec{x},t)}{\left|h^{eq}_{ek}-h^r_e\right|}I_{ek}(\vec{x},t)
  + \frac{h^{eq}_{ik}-h_k(\vec{x},t)}{\left|h^{eq}_{ik}-h^r_e\right|}I_{ik}(\vec{x},t)
\end{equation}
\begin{equation} \label{eq:Liley2}
\left(\frac{\partial}{\partial t} + \gamma_{ek}\right)^2 I_{ek}(\vec{x},t) = 
   e \Gamma_{ek}\gamma_{ek}\left\{N^\beta_{ek}S_e\left[h_e(\vec{x},t)\right]
                               +p_{ek}+\phi_{ek}(\vec{x},t)\right\}
\end{equation}
\begin{equation} \label{eq:Liley3}
\left(\frac{d}{dt} + \gamma_{ik}\right)^2 I_{ik}(\vec{x},t) = 
   e \Gamma_{ek}\gamma_{ek}\left\{N^\beta_{ik}S_i\left[h_i(\vec{x},t)\right]
                               +p_{ik}\right\}
\end{equation}
\begin{equation} \label{eq:Liley4}
\left[\left(\frac{\partial}{\partial t}+v\Lambda\right)^2 
      -\frac{3}{2}v^2\nabla^2\right]\phi_{ek}(\vec{x},t)   = 
N^\alpha_{ek}v^2\Lambda^2 S_e\left[h_e(\vec{x},t)\right]
\end{equation}
\begin{equation} \label{eq:Liley5}
S_k\left[h_k\right] = S^{max}_k\left(
  1 + \exp\left[-\sqrt{2}\frac{h_k-\mu_k}{\sigma_k}\right]
           \right)^{-1}
\end{equation}
\end{subequations}
where index $k=\{e,i\}$ denotes excitatory or inhibitory.  The meaning of the  parameters, along with 
some physiological bounds and the values used in our tests, are given in Table \rf{table:pars}. A detailed description 
of these equations can be found in references \cite{liley2002,frascoli2011}. Here, we will briefly discuss
how the basic electrophysiology is represented. An in-depth discussion of the structure of mean-field
models, and how it reflects observed neurodynamics, can be found in recent reviews by Coombes \cite{coombes2010}
and Bressloff \cite{bressloff2012}.

Equations \rf{eq:Liley1} specify the time derivatives of the mean membrane potentials as originating from
relaxation to their rest value and contributions from the currents associated with each kind of synaptic
connection. Equations \rf{eq:Liley2}-\rf{eq:Liley3} describe the dynamics of the post-synaptic activation,
forced by three input sources. Firstly, there is the input from nearby neurons, which is passed through the
sigmoidal function \rf{eq:Liley5} so that far below threshold $\mu_k$ little or no activation is triggered, while 
far above this threshold, the activation saturates. This kind of nonlinearity is ubiquitous in single
neuron and mean field neuronal models alike. Secondly, there is a constant external input that crudely models
incoming pre-synaptic axonal input from any part of the brain other than the cortex. Finally, there are inputs from 
distant, excitatory neurons, in themselves obeying equations \rf{eq:Liley4}. These are damped wave equations
that describe the spreading of excitatory signals over the cortical surface at finite propagation speed $v$.
\begin{table}[t]
\resizebox{\columnwidth}{!}{%
\begin{tabular}{llllll}
\hline\noalign{\smallskip}
Parameter & Definition & Minimum & Maximum & Value & Units  \\
\hline
$h^r_e$ & resting excitatory membrane potential & $-80$ & $-60$ &-71.3473  & mV\\
$h^r_i$ & resting inhibitory membrane potential & $-80$ & $-60$ & -78.2128 & mV\\
$\tau_e$ & passive excitatory membrane decay time & $5$ & $150$ & 112.891 & ms\\
$\tau_i$ & passive inhibitory membrane decay time & $5$ & $150$ & 116.4642 & ms\\
$h^{\mathrm{eq}}_{ee}$ & excitatory reversal potential & $-20$ & $10$ &6.0551  & mV\\
$h^{\mathrm{eq}}_{ei}$ & excitatory reversal potential & $-20$ & $10$ & -16.8395 & mV\\
$h^{\mathrm{eq}}_{ie}$ & inhibitory reversal potential & $-90$ & $ h^r_k-5$ & -88.0656& mV\\
$h^{\mathrm{eq}}_{ii}$ & inhibitory reversal potential & $-90$ & $ h^r_k-5$ &  -88.6666& mV\\
$\Gamma_{ee}$ & EPSP peak amplitude & $0.1$ & $2.0$ &0.3917 & mV\\
$\Gamma_{ei}$ & EPSP peak amplitude & $0.1$ & $2.0$ & 1.4019 & mV\\ 
$\Gamma_{ie}$ & IPSP peak amplitude & $0.1$ & $2.0$ & 1.4707 & mV\\
$\Gamma_{ii}$ & IPSP peak amplitude & $0.1$ & $2.0$ & 1.4264  &mV\\
$\gamma_{ee}$& EPSP characteristic rate constant$^\ddagger$ & $100$ & $1000$ &551.6  & $\mathrm{s}^{-1}$\\
$\gamma_{ei}$& EPSP characteristic rate constant$^\ddagger$ & $100$ & $1000$ &912.9  & $\mathrm{s}^{-1}$\\
$\gamma_{ie}$& IPSP characteristic rate constant$^\ddagger$ & $10$ & $500$ & 258.5 & $\mathrm{s}^{-1}$\\
$\gamma_{ii}$& IPSP characteristic rate constant$^\ddagger$ & $10$ & $500$ & 96.7 & $\mathrm{s}^{-1}$\\
$N^\alpha_{ee}$ & no.\ of cortico-cortical synapses to excitatory & $2000$ & $5000$ & 4129.3102 & --\\
$N^\alpha_{ei}$ & no.\ of cortico-cortical synapses to inhibitory & $1000$ & $3000$ &4129.3102  & --\\
$N^\beta_{ee}$ & no.\ of excitatory intracortical synapses & $2000$ & $5000$ & 4204.8457 &  --\\
$N^\beta_{ei}$ & no.\ of excitatory intracortical synapses & $2000$ & $5000$ &4204.8457   & --\\
$N^\beta_{ie}$ & no.\ of inhibitory intracortical synapses  & $100$ & $1000$ &987.9069   & --\\
$N^\beta_{ii}$ & no.\ of inhibitory intracortical synapses  & $100$ & $1000$ &210.0476   & --\\
$v$ & axonal conduction velocity & $100$ & $1000$ &251.4 & $\mathrm{cm}\,\mathrm{s}^{-1} $\\
$1/\Lambda$ & decay scale of cortico-cortical connectivity  & $1$ & $10$ &3.6643  & cm \\
$S^{\mathrm{max}}_e$ & maximum excitatory firing rate & $50$ & $500$  &69.4   & $\mathrm{s}^{-1}$ \\
$S^{\mathrm{max}}_i$ & maximum inhibitory firing rate & $50$ & $500$  & 320.9  & $\mathrm{s}^{-1}$ \\
$\mu_e$ & excitatory firing threshold &  $-55$ & $-40$ &-40.9723  & mV \\
$\mu_i$ & inhibitory firing threshold &  $-55$ & $-40$ & -42.5412 & mV \\
$\sigma_e$ & st. deviation of excitatory firing threshold  & $2$ & $7$ & 4.2276  & mV \\
$\sigma_i$ & st. deviation of inhibitory firing threshold  & $2$ & $7$ & 2.1897  & mV \\
$p_{ee}$ & extracortical synaptic input rate & $0$ & $10,000$ &1--10  & $\mathrm{s}^{-1}$ \\
$p_{ei}$ & extracortical synaptic input rate & $0$ & $10,000$ &4.3634  & $\mathrm{s}^{-1}$ \\
\hline
\end{tabular}
}
\caption{Meaning, ranges and values for the model parameters. The values used in the current work were taken from the
data base of pysiologically admissible parameter sets described in Bojak and Liley \cite{bojak2005}.}
\label{table:pars}
\end{table}
We consider the model at the parameter values listed in table \ref{table:pars}. This set of parameter values
was selected from a list of about $80,000$ sets which resulted from an extensive search of parameter space
discussed in Bojak and Liley \cite{bojak2005}. In this search, parameter sets were selected that give rise
to stable equilibrium solutions with physiologically plausible properties. These properties include admissible
values for all variables, a response to noisy input with a strong alpha peak and several other criteria.

In a subsequent paper, the parameter sensitivity of all admissible parameter sets was tested, and it was shown
that the equilibrium turns unstable in a Hopf bifurcation and may, in addition, go through saddle-node
bifurcations \cite{frascoli2011}. The former instability can give rise to complicated spatio-temporal behaviour, while the latter
may cause multistability. Since it is our primary goal to compute time-periodic solutions, we selected
an exemplary parameter set with a supercritical Hopf instability.

Following earlier work on this model, we take the spatial domain to be an $L\times L$ square with periodic boundary 
conditions in
both dimensions. This is a common choice in the study of mean-field models, and is partially justified by the
observation, that each part of the cortex is connected to each other part. A discussion of this argument can be found in
Nunez \& Srinivasan (\cite{Nunez2006}, ch. 11.4). A consequence of this choice, is that the
resulting PDEs are equivariant under reflection in the mid lines and rotation
over ninety degrees, as well as under translations. Equivariance under this rather large symmetry group
greatly complicates the bifurcation scenarios. Solutions that can be created in an equivariant Hopf bifurcation,
for instance, include standing patterns, traveling patterns, and patterns with a different orientation
in two phases of the oscillation \cite{knobloch1992}. The periodic orbit we computed is called a {\em standing square}
in this context.

\section{Numerical treatment}
\label{sec:numerics}
The setup of the simulation and analysis code is described in detail in a previous paper \cite{green2013}.
System \rf{eq:Liley1}-\rf{eq:Liley4} is rewritten as a set of 14 first-order-in-time PDEs, 
which are discretized by finite differences. The discretized system is implemented in PETSc \cite{petsc}, so that
the in-built linear and nonlinear solvers of that software package can be used on the boundary value 
problems that arise in the computation of invariant solutions such as periodic orbits. The solution of
these boundary value problems invariably requires time-stepping of the linearized model, which describes 
the evolution of perturbations to the dependent variables. We have improved the handling of the nonlinear
and linearized models in two ways.

Firstly, the time-stepping of the linearized equations has been implemented as a ``monitor'', i.e. function
that takes input from, but is otherwise independent of, time-stepping of the nonlinear system itself. As
a consequence, we can now use any of the in-built time-stepping algorithms for the nonlinear system.
The results presented below are based on the fourth-order Runge-Kutta scheme. At present, the linearized
equations are still propagated using the implicit Euler scheme, and this may be one reason why we observe a
linear, rather than a quadratic, decrease of the Newton residual of our periodic orbit computation.

Secondly, the boundary value problem for finding periodic orbits is now implemented as a Scalable Nonlinear
Equations Solver (SNES) object. As a consequence, we can use various modified Newton methods implemented
in PETSc. In particular, 
we used the Newton-line search algorithm for the computations below.
The updated code is publicly available from our repository \cite{public}

In the computations below, the system size is fixed to $L=23$(cm), at a resolution of $64\times 64$ grid points,
which yields $57,344$ degrees of freedom. Approximate periodic states were found by forward time stepping,
starting from perturbations from the equilibrium along an unstable eigenvector. We found it hard to obtain
convergent Newton iterations using the boundary value problem with phase conditions to handle the translational
symmetry, as described in Green and van Veen \cite{green2013}. This might be due to the presence of multiple
solutions with different symmetry properties, as predicted by equivariant bifurcation theory \cite{knobloch1992}.
To circumvent this problem, we switched to Dirichlet boundary conditions for the periodic orbit computation.
We set the boundary values according to the bifurcating equilibrium, following the example of Ashwin {\sl et al.}
\cite{ashwin1995}. This step eliminates the translational symmetries, and the primary Hopf bifurcation, which leads 
to a standing square pattern, is
no longer affected by the symmetries of the system. Close to the bifurcation point, the computed solution
approximately satisfies the periodic boundary conditions. Using this setup, we performed Newton--line-search iterations
until the dependent variables have converged up to seven significant
digits. 

\section{The onset of periodic motion}
\label{sec:prelim}
\begin{figure}[t]
\begin{center}
\resizebox{0.75\columnwidth}{!}{\includegraphics{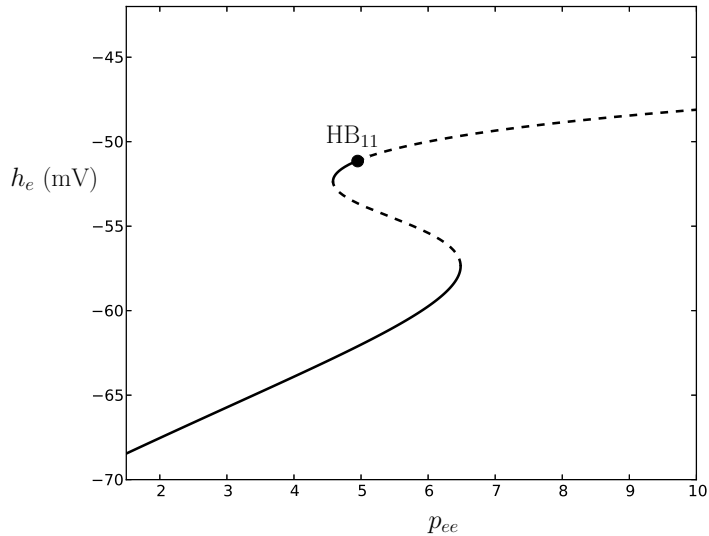} }
\end{center}
\caption{Bifurcation diagram of the spatially homogeneous equilibrium state. Solid lines correspond to stable families
and dashed lines to unstable families. The first instability beyond the saddle-node points is a Hopf bifurcation
with wave numbers $(1,\,1)$. The periodic orbits are computed in between
this point and the second Hopf bifurcation, with wave numbers $(1,\,0)$, shown in figure \ref{fig:eq_pee_detail}.}
\label{fig:eq_pee}
\end{figure}
In figure \ref{fig:eq_pee} a partial bifurcation diagram for the spatially homogeneous equilibrium is shown, constructed
by varying only the external input to the excitatory population. A pair of saddle-node bifurcations causes
multistability in a parameter interval around $p_{ee}=5$. The top part of the family of equilibria is stable up to the Hopf bifurcation.
In fact, this bifurcation is the first of infinitely many, associated with different wave numbers, occurring in a 
small parameter interval. 

\begin{figure}[t]
\begin{center}
\resizebox{\columnwidth}{!}{\includegraphics{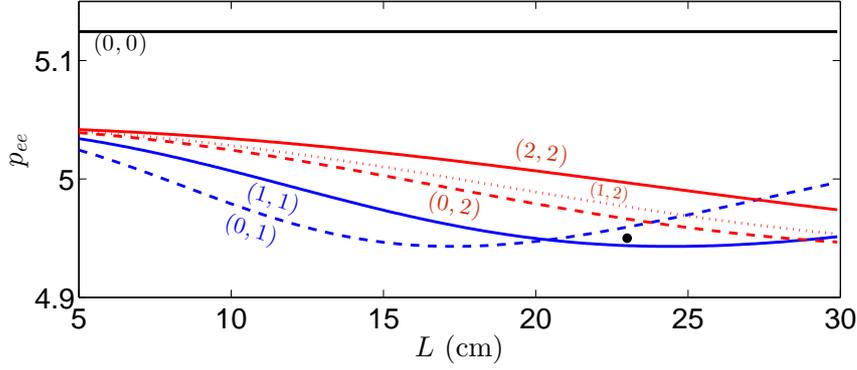} }
\end{center}
\caption{Neutral stability curves for Liley's mean-field model. Shown are Hopf instabilities with
wave numbers up to two (shown between brackets), as the system size $L$ and the external input $p_{ee}$ are varied. All other parameters
are chosen as in table \ref{table:pars}. The periodic orbit shown in figure \ref{fig:linear_snapshots} was
computed at a system size and input parameter represented by the black dot.}
\label{fig:nsc}
\end{figure}
Which wave number turns unstable first is determined by the system size, as can be seen from the neutral stability
curves in figure \ref{fig:nsc}. Each of the curves shown divides the two parameter plane into regions where
perturbations with the specified wave numbers grow or decay. Below the union of these curves, only stable equilibrium
solutions are expected to exist. As the system size increases, the wave numbers of the leading instability increase as well,
while the length scale of the leading instability saturates at about $16$(cm), and its period around $100$(ms),
in the alpha range. At 
our current system size, $L=23\,\text{(cm)}$, the primary instability has one sinusoidal oscillation
in both spatial directions, i.e. it has wave numbers $(1,\,1)$ and thus satisfies both periodic and homogeneous Dirichlet
boundary conditions.  

\begin{figure}[t]
\begin{center}
\resizebox{0.85\columnwidth}{!}{\includegraphics{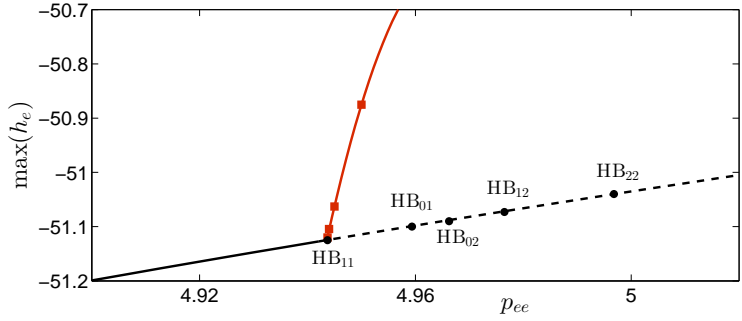} }
\end{center}
\caption{Detail of the bifurcation diagram around the first Hopf bifurcation, on the top branch of figure \ref{fig:eq_pee}.
The solid, black curve corresponds to a stable equilibrium, the dashed black curve to an unstable equilibrium and 
the red curve to a stable, spatially inhomogeneous periodic orbit. For the latter, the maximum of $h_e$ at $(x,\,y)=(L/4,\,L/4)$ is
plotted. Red squares correspond to computed periodic orbits and the red line has been added for visibility. Black dots denote Hopf bifurcations for wave numbers
up to two, indicated by the subscripts.}
\label{fig:eq_pee_detail}
\end{figure}
\begin{figure}[t]
\begin{center}
\resizebox{0.85\columnwidth}{!}{\includegraphics{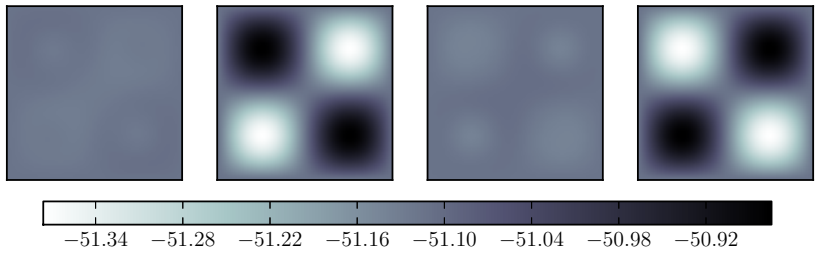} }
\end{center}
\caption{Snapshots of the standing square solution at $p_{ee}=4.95$. Displayed is the excitatory 
membrane potential ($h_e$) at 24.6 \textrm{ms} intervals.  
The grey scale is in \textrm{mV}.}
\label{fig:linear_snapshots}
\end{figure}

 
When increasing the forcing, $p_{ee}$, at this fixed system size, we cross infinitely many Hopf bifurcations with increasing wave
numbers, accumulating at $p_{ee}\approx 5.04$. Those with wave numbers up to two are shown in the detailed bifurcation 
diagram \ref{fig:eq_pee_detail}. The red curve in this diagram denotes the standing square periodic solution that emanates
from the leading Hopf bifurcation. Snapshots of this solution are shown in figure \ref{fig:linear_snapshots}. This close
to the bifurcation point, the pattern still looks very much like the sinusoidal eigenmode, and has a period of $98$(ms).

\begin{figure}[t]
\begin{center}
\resizebox{0.85\columnwidth}{!}{\includegraphics{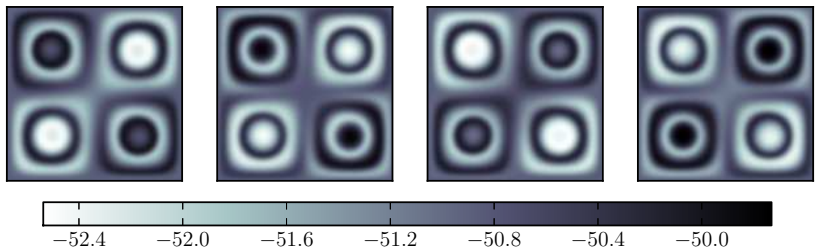} }
\end{center}
\caption{Snap shots the excitatory membrane potential, perturbed away from the equilibrium
at $p_{ee}=5.1$ along the unstable eigenmode with wave numbers $(1,\,1)$. The snapshots are
$24$(ms) apart. In this transient motion, the frequency of the oscillation at the centre
of the four $L/4 \times L/4$ squares increases, while rings form around them.}
\label{fig:nonlinear_snapshots}
\end{figure}

If we increase the forcing beyond the accumulation point of Hopf bifurcations, complex spatio-temporal dynamics is observed.
Figure \ref{fig:nonlinear_snapshots} shows a transient, produced by perturbing the unstable equilibrium along the
unstable eigenvector involved in the leading stability. Rings are forming around the four extrema of the sinusoidal  pattern,
corresponding to a difference in the phase of the oscillation. At the same time, the frequency of the oscillation at
the centres increases. We conjecture, that this process will lead to the formation of ``hot spots'' seen in the onset
of gamma band activity in this model \cite{bojak2007}. In ongoing work, we attempt to compute periodic solutions 
that exhibit hot spots. The standing square orbit may well show the initial stage of spot formation, but we may need
to track the the attracting solution through a sequence of bifurcations, in which the behaviour becomes increasingly localized.

\section{Conclusion}

We report here on spatially inhomogeneous, time-periodic solutions to a full-fledged, PDE-based 
mean-field cortex model. To the best of our knowledge, ours is the first attempt to compute
such nonlinear invariant solutions for this class of models. The computation is challenging because of
the large number of degrees of freedom in the discretized system, the strong nonlinearities
and the symmetries of the system that complicate the bifurcation diagram. 

This is only the first step in uncovering the dynamical repertoire of the PDE-based model.
In ongoing work, we will continue the periodic orbit in the external
forcing, and examine its role in the transition from smooth oscillations to localized hot spots
and travelling waves. There is mounting evidence, based on Voltage 
Sensitive Dye (VSD) experiments, that such coherent structures pay a role in information
processing (e.g. \cite{muller2013}). 

A combination of experimental work and predictive modelling should help us gain insight in the
role the cortex plays in
high-level brain functioning, as well as in the generation of pathological states such as spreading depression and
epileptic seizures. Thus, through this work we hope to contribute indirectly to the 
better understanding and, hopefully, future treatment of, these common diseases. This noble goal
attracts many researchers from different areas of research, not in the last place Hilda Cerdeira, to whose
live and work this 
special issue is devoted.

\end{document}